\documentclass{llncs} 
\pdfoutput=1

\usepackage[utf8]{inputenc}
\usepackage{float}
\usepackage{graphicx}
\usepackage{amsmath}
\usepackage{hyperref}

\begin{document}

%
%

\title{Privacy by Design: \\ From  Technologies to Architectures}
\subtitle{(Position Paper)\protect\footnote{The final publication is available at link.springer.com (\url{http://link.springer.com/chapter/10.1007/978-3-319-06749-0_1}).}}

\author{Thibaud Antignac \and Daniel Le M\'etayer}

\institute{Inria, Universit\'e de Lyon\\
\email{thibaud.antignac@inria.fr/daniel.le-metayer@inria.fr}}


\maketitle

%
%

\begin{abstract}
Existing work on privacy by design mostly focus on technologies rather than methodologies and on components rather than architectures. In this paper, we advocate the idea that privacy by design should also be addressed at the architectural level and be associated with suitable methodologies. Among other benefits, architectural descriptions enable a more systematic exploration of the design space. In addition, because privacy is intrinsically a complex notion that can be in tension with other requirements, we believe that formal methods should play a key role in this area. After  presenting our position, we provide some hints on how our approach can turn into practice based on ongoing work on a privacy by design environment.
\end{abstract}

%
%

\section{Introduction}

Privacy by design is often held up as a necessary step to improve privacy protection in the digital society~\cite{Langheinrich,Poullet}. It will even become a legal obligation in the European Community\footnote{And also ``a prerequisite for public procurement tenders according to the Directive of the European Parliament and of the Council on public procurement as well as according to the Directive of the European Parliament and of the Council on procurement by entities operating in the water, energy, transport and postal services sector.''} if the current draft of Data Protection Regulation~\cite{ecregulation2013} eventually gets adopted. The fact that future regulatory frameworks promote or impose privacy by design is definitely a positive step but their adoption will take time, they are unlikely to be applicable in all regions of the world and  the interpretation of the principle itself may vary greatly\footnote{And it is obviously out of question to define it very precisely in a legal document.} among jurisdictions. Therefore, the possible adoption of privacy by design laws should obviously not be seen as the end of the story: the extent to which a true privacy by design will actually turn to reality will also depend on other factors such as social demand, economic conditions and of course the availability of technical solutions. Even though all these dimensions are essential and all possible means should be used to foster the adoption of privacy by design, we choose to focus on the technical issues in this paper. 

As discussed by several authors~\cite{diaz,hoepman2013,kerschbaum2012,Hafiz}, a range of  privacy enhancing technologies (PETs) are now available, which can provide strong privacy guarantees in a variety of contexts. Even if more research on PETs will always be needed, a question that arises at this stage is how to favor the uptake of these technologies. The position that we want to promote and bring forward for discussion in this paper is threefold:
\begin{enumerate}
\item Existing work in this area mostly focus on technologies rather than methodologies and on components rather than architectures. We advocate the idea that privacy by design should also be addressed at the architectural level and be associated with suitable methodologies. Among other benefits, architectural descriptions enable a more systematic exploration of the design space.  
\item Because privacy is intrinsically a complex notion, which, in addition, often turns out to be (or seems to be) in tension with other requirements, formal methods should play a key role in this area. Among other benefits, formal descriptions make it possible to define precisely the concepts at hand and the guarantees provided by a solution, to reason about the design choices  and ultimately to justify them rigorously, hence contributing also to accountability.
\item Because designers are generally not experts in formal methods, dedicated frameworks and user-friendly interfaces should be designed to hide the complexities of the models and make them usable in this context. Such frameworks should  allow  designers to express  their requirements and to interact with the system to refine or adapt them until a satisfactory architecture is obtained.
\end{enumerate}

The paper is organized as follows: We present our position and discuss it in more detail in Section 2. In order to illustrate this position, we proceed in Section 3 with an example of formal privacy logic which is applied in Section 4 to the design of smart metering systems. Section 5 discusses related work and Section 6 outlines directions for further research.

\section{Position}

A wide array of techniques have been proposed or applied to the definition of new PETs during the last decades~\cite{Deswarte,Goldberg,kerschbaum2012,Rezgui}, including zero-knowledge proofs, secure multi-party computation, homomorphic encryption, commitments, private information retrieval (PIR), anonymous credentials, anonymous communication channels, or trusted platform modules (TPM). These PETs have been used to provide strong privacy guarantees in a variety of contexts such as
smart metering~\cite{Garcia,LeMay:07,Rial}, electronic traffic pricing~\cite{Troncoso,Jacobs,Popa}), ubiquitous computing~\cite{Langheinrich} or location based services~\cite{Damiani:10,Kosta,Krumm}. Even if more techniques will always be needed to defeat new attacks in this eternal game between cops and robbers,  there is now a need for serious thinking about the conditions for the adoption of existing PETs by industry. This reflection has started with convincing cases for the development of appropriate methodologies~\cite{diaz,hoepman2013,Wing} or development tools~\cite{kerschbaum2012} for privacy by design. In this paper, we pursue this line of thought in arguing that privacy by design should also (and firstly) be considered at the architecture level and should be supported by appropriate reasoning tools relying on formal models. 

Let us consider the case for architectures first. Many definitions of architectures have been proposed in the literature. In this paper, we will adopt a definition inspired by~\cite{bass2013}\footnote{This definition is a generalization (to system architectures) of the definition of software architectures proposed in~\cite{bass2013}.}: \textit{The architecture of a system is the set of structures needed to reason about the system, which comprise software and hardware elements, relations among them and properties of both.} The atomic components of an architecture are coarse-grain entities such as modules, components or connectors. 
In the context of privacy, as suggested in Section 3, the components are typically the PETs themselves and the purpose of the architecture is their combination to achieve the privacy requirements of the system.
Therefore, an architecture is foremost an abstraction and ``this abstraction is essential to taming the complexity of a system -- we simply cannot, and do not want to, deal with all the complexity all the time''~\cite{bass2013}. 

Most of the reasons identified in~\cite{bass2013} to explain why architectures matter are very relevant for privacy by design:
\begin{itemize}
\item The architecture is a carrier of the earliest and hence most fundamental hardest-to-change design decisions: overlooking architectural choices may thus seriously hamper the integration of privacy requirements in the design of a system. 
\item Architectures channel the creativity of developers, reducing design and system complexity: because they make it possible to abstract away unnecessary details and to focus on critical issues, architectures can help privacy designers reason about privacy requirements and the combination of PETs which could be used to meet them.
\item A documented architecture enhances communication among stakeholders: documenting the design choices is especially important in the context of privacy to meet the obligations arising from Privacy Impact Assessments (PIA) and accountability (which are emphasized in the 
Data Protection Regulation draft~\cite{ecregulation2013}).
\item An architecture can be created as a transferable, reusable model: architectures can therefore play a key role to increase the reusability of privacy friendly solutions, which could lead to significant cost reductions and better dissemination of knowledge and expertise among developers. Being able to go beyond case-by-case strategies and to factorize efforts is a key step for privacy by design to move from craft to industry.  
\end{itemize}

If the choice is made to work at the architectural level, the next question to be answered is: ``how to define, represent and use architectures ?'' In practice, architectures are often described in a pictorial way, using different kinds of graphs with legends defining the meaning of nodes and vertices, or semi-formal representations such as UML diagrams (class diagrams, use case diagrams, sequence diagrams, communication diagrams, etc.). The second point we want to make here is that, even though such pictorial representations can be very useful (especially when their use is standardized), reasoning about privacy requirements is such a subtle and complex issue that the architecture language used for this purpose must be defined in a formal way. By formal, we mean that the properties of the architectures must be defined in a mathematical logic and reasoning about these properties must be supported by a formal proof or verification system.  Reasoning about privacy is complex for different reasons: first, it is a multi-faceted notion stemming from a variety of principles which are not defined very precisely. It is therefore of prime importance to circumscribe the problem, to define precisely the aspects of privacy which are taken into account and how they are interpreted in the design choices. The fact that not all aspects of privacy are susceptible to formalization is not a daunting obstacle to the use of formal methods in this context: the key point is to build appropriate models for the aspects of privacy (such as data minimization) which are prone to formalization and involve complex reasoning. Another source of complexity in this context is the fact that privacy often  seems to   conflict with other requirements such as guarantees of authenticity or correctness, efficiency, usability, etc. 
To summarize, formal methods should:
\begin{itemize}
\item Make it possible to define precisely the concepts at hand (requirements, assumptions, guarantees, etc.).
\item Help designers to explore the design space and to reason about the possible choices. Indeed, privacy by design is often a matter of choice~\cite{lemetayer2010}: multiple options are generally available to achieve a given set of functionalities, some of them being privacy friendly, others less, and a major challenge for  the designer is to understand all these options, their strengths and weaknesses.
\item Provide a documented and rigorous justification of the design choices, which would be a significant contribution  to the accountability requirement. Accountability is defined in Article 22 of the current draft of the future Data Protection Regulation~\cite{ecregulation2013} as the following obligation for data collectors: ``The controller shall adopt appropriate policies and implement appropriate and
demonstrable technical and organizational measures to ensure and be able to
demonstrate in a transparent manner that the processing of personal data is
performed in compliance with this Regulation...''
\end{itemize}

It should be clear however, that designers are generally not experts in formal methods and appropriate tools and user-friendly interfaces have to be made available to hide the complexities of the models. Such frameworks should  allow  designers to express  their requirements, to get understandable representations (or views) of the architectural options,  to navigate in these views, and to refine or adapt them until a satisfactory architecture is obtained.

Considering all these requirements, it should be clear that specific, dedicated formal frameworks have to be devised to support privacy by design. To make the discussion more concrete, we provide in the following sections some hints about such a framework and its application to a real case study.

\section{Example of Formal Model}

As an illustration of the position advocated in the previous section, we present now a subset of a privacy logic to reason about two key properties: the minimization (of the collection of personal data) and the accuracy (of the computations) . Indeed, the tension between data minimization and accuracy is one of the delicate issues to be solved in many systems involving personal data. For example, electronic traffic payment systems~\cite{Troncoso,Jacobs,Popa} have to guarantee both the correctness of the computation of the fee and the limitation of the collection of location data; smart metering systems~\cite{Garcia,LeMay:07,Rial} have also to ensure the  correct computations of the fees and to limit  the collection of consumption data; recommendation systems~\cite{McSherry:2009} need to achieve both the accuracy of the recommendations while minimizing the disclosure of personal preferences; electric vehicle charging systems~\cite{Kargl} also need to guarantee the accuracy of the bills and the protection of the location information of the vehicles.

\subsection{Excerpt of a Privacy Epistemic Logic}

Because privacy is closely connected with the notion of knowledge, epistemic logics form an ideal basis to reason about privacy properties. Epistemic logics~\cite{LivreHalpern} are a family of modal logics using a knowledge modality usually denoted by $K_i\left(\phi\right)$ to express the fact that agent $i$ knows the property $\phi$. However standard epistemic logics based on possible worlds semantics suffer from a weakness which makes them unsuitable in the context of privacy: this problem is often referred to as ``logical omniscience''~\cite{HalpernPucellaLO}. It stems from the fact that agents know all the logical consequences of their knowledge (because these consequences hold in all possible worlds). An undesirable outcome of logical omniscience would be that, for example, an agent knowing the commitment $C\left(v\right)$ (or hash) of a value $v$ would also know $v$ (or the possible values of $v$). This is obviously not the intent in a formal model of privacy where commitments are precisely used to hide the original values to the recipients. This issue is related to the fact that standard epistemic logics do not account for limitations of computational power.

Therefore it is necessary to define dedicated epistemic logics to deal with different aspects of privacy and to model the variety of notions and techniques at hand (e.g. knowledge, zero-knowledge proof, trust, etc.). 
Let us consider for the sake of illustration the following excerpt of a privacy epistemic logic:

\begin{align}
    \phi ::= \, &\phi_0\\
        | \, &\neg \phi\\
        | \, &\phi \wedge \phi^{\prime}\\
        | \, &\text{K}_i \left( \phi_0 \right)\\
        | \, &\text{X}_i \left( \phi_0 \right)\\
\nonumber \\
    \phi_0 ::= \, &\text{receive}_{i,j} \left(x \right)\\
                 \mid\, &\text{receive}_{i,j} \left( \textit{prim} \right)\\
             \mid\, &\text{trust}_{i,j}\\
             \mid\, &\text{compute}_i \left( x = t \right) \\
             \mid\, &\text{check}_i \left( \textit{eq} \right)\\
             \mid\, &\text{has}_i \left( x \right)\\
             \mid\, &\text{prim} \mid \textit{p} \mid \phi_0 \wedge \phi_0^{\prime} \\
\nonumber \\
    \textit{prim} ::= \, &\text{proof}_{i,j} \left( \textit{p} \right) \mid \textit{att} \mid \textit{prim} \wedge \textit{prim}^{\prime}\\
    \textit{p} ::= \, &\textit{att} \mid \textit{eq} \mid p \wedge p^{\prime}\\
    \textit{att} ::= \, &\text{attest}_i \left( \textit{eq} \right)\\
    \textit{eq} ::= \, &\textit{t}\; \textit{rel}\; \textit{t}^{\prime}\\
    \textit{rel} ::= \, &= \,\mid\, < \,\mid\, > \,\mid\, \leq \,\mid\, \geq\\
    \textit{t} ::= \, &c \mid x \mid \text{F} \left( \textit{t}_1, \dots, \textit{t}_n \right)
\end{align}

This logic involves only two modalities: the usual knowledge operator $K_i$ and the ``algorithmic knowledge''~\cite{LivreHalpern,Pucella} denoted by $X_i$, which represents the knowledge that an agent $i$ can actually build.
Properties $\phi_0$ include both the basic properties $p$ on variables of the system and properties used to characterize the architecture itself. More precisely, $\text{receive}_{i,j} \left( x \right)$ means that agent $j$ can send the variable $x$ to agent $i$, $\text{receive}_{i,j} \left( \textit{prim} \right)$ means that agent $j$ can send the property $prim$ to agent $i$ where $prim$ can be a combination of (non interactive) zero-knowledge proofs and attestations. An attestation $\text{attest}_i \left( \textit{eq} \right)$ is a simple declaration by agent $i$ that property $\textit{eq}$ holds. This declaration is of no use to agent $j$ unless $j$ trusts $i$, which is expressed by $\text{trust}_{j,i}$. The primitive $\text{proof}_{i,j} \left( \textit{p} \right) $ means that agent $i$ can make a proof of property $p$ which can be checked by agent $j$\footnote{In general a proof could be checked by several agents, which could be expressed using a set of agents names as second index.}. The properties $\text{compute}_i \left( x = t \right)$ and $\text{check}_i \left( p \right)$ are used to express the fact that an agent $i$ can respectively compute a variable $x$ defined by the equation $x = t$ or check a property $p$. Symbol $F$ stands for the available basic operations (e.g. hash, homomorphic hash, encryption, random value generation, arithmetic operations, etc.), $c$ stands for constants, and $x$ for variables. Last but not least, $\text{has}_i \left( x \right)$ is the property that agent $i$ can get the value of variable $x$ (which does not, in itself, provide any guarantee about the correctness of this value).

\subsection{Semantics and Axiomatization}

The semantics of this logic can be defined using an augmented Kripke model $M = \left( \textit{Arch}, \pi, \mathcal{D}_1, \dots, \mathcal{D}_n \right)$ where:

\begin{itemize}
    \item $\textit{Arch}$ is a set of possible architectures (generally called worlds) of the system under consideration. In our setting, an architecture is defined as a property  $\phi_0$ which characterizes all the operations available to the agents.
    \item $\pi$ is an interpretation for the primitives $\textit{prim}$ and the relations $\textit{eq}$.
    \item $\mathcal{D}_1, \dots, \mathcal{D}_n$ are the deductive systems associated with agents $1, \ldots, n$.
    
\end{itemize}  
 
The deductive system associated with an agent $i$ makes it possible to define the semantics of the $X_i$ operator (knowledge built by agent $i$). In other words, each agent $i$ can apply rules $\triangleright_i$ to derive new knowledge. Typical rules of deductive systems include:

  \begin{itemize}
            \item  $\text{receive}_{i,j} \left( \textit{prim} \right) \triangleright_i \textit{prim}$
            \item  $\text{attest}_j \left( \textit{eq} \right), \text{trust}_{i,j}  \triangleright_i \textit{eq}$
            \item  $\text{proof}_{j,i} \left( \textit{p} \right) \triangleright_i \textit{p}$
            \item  $\text{check}_i \left( \textit{eq} \right) \triangleright_i \textit{p}$
            \item  $\text{compute}_i \left( x = t \right) \triangleright_i x = t$
            \item  $\text{hash} \left( x_1 \right) = \text{hash} \left( x_2 \right) \triangleright_i x_1 = x_2$
            \item $\text{hhash} \left( x \right) = \text{hhash} \left( x_1 \right) \otimes \text{hhash} \left( x_2 \right) \triangleright_i x = x_1 + x_2$
            \item $\text{receive}_{i, j} \left( x \right) \triangleright_i \text{has}_i \left( x \right)$
            \item $\text{compute}_i \left( x = t \right) \triangleright_i \text{has}_i \left( x \right)$
            \item $\text{has}_i \left( x_1 \right), \dots, \text{has}_i \left( x_m \right), \text{dep} \left( x, \left\{ x_1, \dots, x_m \right\} \right) \triangleright_i \text{has}_i \left( x \right)$ \\ \hspace{\parindent} with $\text{dep} \left( x, \left\{ x_1, \dots, x_m \right\} \right)$ a dependence relationship known by $i$ stating that $x$ can be derived from $x_1$, \dots,  $x_m$.
\end{itemize}

In order to reason about architectures, we can use an axiomatization in the style of previous work on deductive algorithmic knowledge~\cite{LivreHalpern,Pucella}. Typical examples of axioms and inference rules include: 
\begin{align}
    \textbf{TAUT}\text{:}\; &\text{All instances of propositional tautologies.}\\
    \textbf{MP}\text{:}\; &\text{From $\phi$ and $\phi \rightarrow \psi$ infer $\psi$}\\
    \textbf{Gen}\text{:}\; &\text{From $\phi$ infer $\text{K}_i \left(\phi\right)$}\\
    \textbf{K}\text{:}\; &\text{K}_i \left( \phi \rightarrow \psi \right) \rightarrow \left( \text{K}_i \left(\phi\right) \rightarrow \text{K}_i \left(\psi\right) \right)\\
    \textbf{T}\text{:}\; &\text{K}_i \left(\phi\right) \rightarrow \phi\\
    \textbf{KC}\text{:}\; &\text{K}_i \left( \phi \wedge \psi \right) \rightarrow \left( \text{K}_i \left(\phi\right) \wedge \text{K}_i \left(\psi\right) \right) \\
    \textbf{XD}\text{:}\; &\text{From $X_i(\phi_1)$, \ldots, $X_i(\phi_n)$ and $\phi_1,\ldots, \phi_n \triangleright_i \phi $  infer $X_i(\phi)$} \\
    \textbf{XT}\text{:}\; &\text{X}_i \left(\phi\right) \rightarrow \phi\\
    \textbf{XC}\text{:}\; &\text{X}_i \left( \phi \wedge \psi \right) \rightarrow \left( \text{X}_i \left(\phi\right) \wedge \text{X}_i \left(\psi\right) \right)
\end{align}

A key difference between the properties of $K_i$ and $X_i$ is the lack of counterpart of axioms  $\textbf{Gen}$ and $\textbf{K}$ for $X_i$, which  makes it possible to avoid the omniscience problem. In contrast, the knowledge built by an agent $i$ depends on the associated deductive system $\triangleright_i$ as expressed by rule $\textbf{XD}$. Rules $\textbf{T}$ and $\textbf{XT}$ express the fact that an agent cannot derive false properties.

The axiomatization outlined in this section forms the basis for an inference algorithm which makes it possible to prove properties of architectures (expressed as $\phi_0$ properties) as discussed in the next subsection.

\subsection{Use of the Formal Model}
\label{sec:use-of-the-formal-model}

The first functionality of a design environment is to make it possible for the designer to express the requirements that apply to the system and, optionally, the design choices which have already been made (or which are imposed by the environment or the client) as well as the possible trust assumptions. 

Formally speaking, the requirements are made of three components:
\begin{itemize}
 \item Functional requirements: the purpose of the system, which is expressed as a set of equations $x = t$.
 \item Privacy and knowledge requirements: values that should not (respectively should) be known by certain actors, which is expressed by properties $ \neg \text{has}_i \left( x \right)$ (respectively $\text{has}_i \left( x \right)$).
 \item Correctness (integrity) requirements: the possibility for certain actors to ensure that certain values are correct, which is expressed as $\text{X}_i \left( \textit{eq} \right)$.
\end{itemize}
Generally speaking, other non-functional requirements could also be considered but the combination of privacy and correctness already provides a degree of complexity which is sufficient to illustrate the approach.

The design choices, which add further requirements on the architecture, can be defined as $\phi_0$ properties. They can express, for example, the fact that communication links are (or are not) available between specific components or certain computations (zero-knowledge, homomorphic hash, etc.) can (or cannot) be performed by certain components.

Several situations are possible when the designer has entered this first batch of information:
\begin{enumerate}
 \item The requirements may be contradictory, for example because privacy requirements conflict with knowledge or architectural requirements or because architectural requirements themselves are not consistent (operations computed by components which cannot get access to the necessary parameters, checks which cannot be carried out because some values are not available to the component, etc.). The system returns the identified contradictions, which may provide some hints to the designer to modify his initial input.
 \item The requirements may be consistent but not precise enough to characterize a unique architecture. In this case, the system can use a library of existing PETs to provide suggestions to the user. The user can then decide to apply a given PET (which is expressed formally by the addition of a new assumption, e.g. $\text{receive}_{i,j} \left( \text{proof}_{j,i} \left( \textit{p} \right) \right) $ for a zero-knowledge proof of property $p$ sent by agent $j$ to agent $i$. 
  \item The requirements may be precise enough to specify a unique (and correct architecture).
\end{enumerate}

The first two cases lead to a new iteration of the procedure. In the last case, the designer has obtained a satisfactory architecture (which does not prevent him from performing a new iteration with different assumptions to further explore the design space).

\section{Example of Application}
\label{sec:example-of-application}

Let us now illustrate the framework outlined in the previous section with a small smart metering case study. Privacy friendly smart grids and smart metering systems have been studied extensively \cite{Garcia,Kerschbaum,Jawurek,Rial}. Our purpose here is neither to present a new solution nor to provide a comprehensive study of smart-metering systems but to show how a formal framework can help a designer to find his way among the possible options. We focus here on the billing functionality and the potential tensions between the privacy requirements of the clients and the need for the operator to ensure the correctness of the computation of the fee.
 
Let us assume in the first instance that the system is made of three components: the back-end system of the operator $o$, the computer of the user (customer) $u$ and the meter $m$. The requirements for this case study are the following:
\begin{itemize}
 \item Functional requirements: the purpose of the system is the computation of the fee which is expressed as the equation $Fee = \sum_{i=1}^n(P(C_i))$ where $C_i$ are the actual consumption values for the considered period of time ($i\in[1,n]$) and $P$ is the cost function.
 \item Privacy and knowledge requirements:   $ \neg \text{has}_o \left( C_i \right)$ and  $\text{has}_o \left( Fee \right)$ respectively, to express the fact that the operator $o$ should not obtain the individual consumption values $C_i$ but should get the fee.
 \item Correctness (integrity) requirements: the operator must be sure that the fee is correct: $\text{X}_o \left(Fee = \sum_{i=1}^n(P(C_i)) \right)$.
\end{itemize}

Let us assume in a first scenario that the designer considered a direct communication link between the meter and the operator, which means that the architecture would include the property $\text{receive}_{o,m} \left(C_i \right)$. This possibility would obviously conflict with the privacy requirement since $\text{receive}_{o, m} \left( C_i \right) \triangleright_i \text{has}_o \left( C_i  \right)$. Two communication links are therefore necessary: from $m$ to $u$ and from $u$ to $o$. 

The next question is where the computation of $P$ should take place: generally speaking, this could be at the back-office of the operator, on the meter, or on the computer of the user. 
Depending on his initial ideas about the architecture and the constraints imposed by the hardware, the designer can either enter directly the appropriate $compute_o$, $compute_m$ or $compute_u$ property. Otherwise these options would be suggested in turn by the system.

\begin{itemize}
\item
The first option turns out to conflict with the privacy requirements (because the operator would need the input values $C_i$ to compute $Fee$) unless homomorphic encryption can be used to allow the operator to compute $Fee$ on encrypted values.
 \item 
The second option can be further explored if it does not incur inacceptable additional costs on the meters. However, the system would then identify a trust requirement (which may or may not have been foreseen by the designer): the operator must trust the meter ($trust_{o,m}$) because the only information received by the operator would be an attestation of the meter  ($attest_m \left(Fee = \sum_{i=1}^n\left(P\left(C_i\right)\right) \right)$) and an attestation can turn into a true guarantee only in conjunction with a trust assumption (as expressed by the rule $\text{attest}_j \left( eq \right), trust_{i,j} \triangleright_i eq$).  
\item
The third option will be more appealing  if the role of the meters has to be confined to the delivery of the consumption measures. But this choice leads to another requirement: either the user can be trusted by the operator (which is excluded by assumption) or he has to be able to provide to the operator a proof of correctness of the computation of the fee~\cite{Rial} ($receive_{o,u} \left( \text{proof}_{u,o} \left( Fee = \sum_{i=1}^n\left(P\left(C_i\right)\right) \right) \right)$). 
\end{itemize}

If none of these options is available, further actors need to be considered and involved in the computation of the fee. In general, these actors can either be pairs or trusted third parties. In both cases, further requirements would arise: a secure multi-party computation scheme would be necessary to ensure that the pairs do not learn each other's consumptions and trust assumptions would be necessary to ensure that computations can be delegated to the third party.   

As discussed in Section 2, designers are usually not experts in formal methods. As a result, a design environment should provide alternative modes of interaction hiding the complexities of the formal model. Considering that designers are used to graphical notations, we have chosen to represent the different views of the architecture as annotated graphs.  The interested reader can find in Annex~1 the ``location views'' showing, for the two successful scenarios described above, the architectures obtained at the end of the interaction process.

\section{Related Work}

This position paper stands at the crossroads of at least three different areas: software architectures, formal models for privacy and engineering of privacy by design.  

Software architectures have been an active research topic for several decades \cite{shaw2006} but they are usually defined using purely graphical, informal means \cite{bass2013} or within semi-formal \cite{booch2013}   frameworks. Formal frameworks have been proposed to define software architectures \cite{allan1994,inverardi1995,lemetayer1996,perry1992} but they are usually based on process algebras or graph grammars and they are not designed to express privacy properties. One exception is the framework introduced in \cite{lemetayer2013} which defines the meaning of the available operations in a (trace-based) operational semantics and proposes an inference system to derive properties of the architectures. The inference system is  applied to the proof of properties related to the use of spot checks in electronic traffic pricing systems. Even though the goal of \cite{lemetayer2013} is to deal with architectures, it remains at a lower level of abstraction than the framework sketched here (because of its operational underpinning, which contrasts with the epistemic logic used here) and can hardly be extended to other privacy mechanisms. 

On the other hand, dedicated languages have been proposed to specify privacy properties ~\cite{Backes:04,Barth:06,Becker:10,Fong:11,DLM:09,Li:06,May:06,Yu:04} but the policies expressed in these languages are usually more fine-grained than the properties considered here because they are not intended to be used at the architectural level.
Similarly, process calculi such as the applied pi-calculus~\cite{abadi2001} have been applied to define privacy protocols~\cite{delaune2009}. Because process calculi are general frameworks to model concurrent systems, they are more powerful than  dedicated frameworks.
The downside is that protocols in these languages are expressed at a lower level and the tasks of specifying a protocol and its expected properties are more complex. Again, the main departure of the approach advocated in this paper with respect to this trend of work is that we reason at the level of architectures, providing ways to express properties without entering into the details of specific protocols.

Notions such as \emph{k}-anonymity~\cite{Li:11,Sweeney}, $l$-diversity~\cite{Kifer:06} or $\epsilon$-differential privacy~\cite{Dwork:06,Dwork:11} have also been proposed as ways to measure the level of privacy provided by an algorithm. 
Differential privacy provides strong privacy guarantees independently of the background knowledge of the adversary: the main idea behind $\epsilon$-differential privacy  is that the presence (or the absence) of an item in a database (such as the record of a particular individual) should not  change in a significant way the probability of obtaining a certain answer for a given query. 
Methods~\cite{Dwork:11,McSherry:10,McSherry:07} have been proposed to design algorithms achieving privacy metrics or to verify that a system achieves a given level of privacy~\cite{Kaynar}. These contributions on privacy metrics  are complementary to the work described in this paper. We follow a logical (or qualitative) approach here, proving that a given privacy property is met (or not) by an architecture. As suggested in the next section, an avenue for further research would be to cope with quantitative reasoning as well, using inference systems to derive properties expressed in terms of privacy metrics.  
 
As far as the engineering of privacy is concerned, several authors \cite{diaz,hoepman2013,kerschbaum2012,mulligan2012,Spiekermann} have already pointed out the complexity of problem as well as the ``richness of the data space''\cite{diaz}, calling for the development of more general and systematic methodologies for privacy by design. \cite{hoepman2013} has used design patterns to define eight privacy strategies\footnote{Strategies are defined as follows in \cite{hoepman2013}: ``A design strategy describes a fundamental approach to achieve a certain design goal. It has certain properties that allow it to be distinguished from other (fundamental) approaches that achieve the same goal.''} called respectively: Minimise, Hide, Separate, Aggregate, Inform, Control, Enforce and Demonstrate. As far as privacy mechanisms are concerned, \cite{kerschbaum2012} points out the complexity of their implementation  and the large number of options that designers have to face. To address this issue and favor the adoption of these tools, \cite{kerschbaum2012} proposes a number of guidelines for the design of compilers for secure computation and zero-knowledge proofs. In a different context (designing information systems for the cloud), \cite{manousakis2013} also proposes  implementation techniques to make it easier for developers to take into account privacy and security requirements. In the same spirit, \cite{Pearson} proposes a decision support tool based on design patterns to help software engineers to take into account privacy guidelines in the early stage of development.

A recent proposal (\cite{PEARs}) also points out the importance of architectures for privacy by design. \cite{PEARs} proposes a design methodology for privacy (inspired by \cite{bass2013}) based on tactics for privacy quality attributes (such as minimization, enforcement or accountability) and privacy patterns (such as data confinement, isolation or Hippocratic management). The work described in \cite{PEARs} is complementary to the approach presented here: \cite{PEARs} does not consider formal  aspects while this paper does not address the tactics for privacy by design.

Finally, even though we decided to focus on one (important) aspect of privacy by design here, namely data minimization, other aspects also deserve more attention, in particular the design of appropriate interfaces to allow users to take more informed decisions about their personal data \cite{mulligan2012}.

\section{Directions for further work}

The basic idea put forward in this paper is that  privacy by design should be addressed at the architecture level and this should be done, at least for critical aspects such as minimization, in a formal way. As stated in Section 1, some aspects of privacy (such as proportionality or purpose) may be more difficult to formalize or impossible to formalize entirely. This should not be an obstacle to the use of formal methods for other aspects of privacy. Generally speaking, architectures are made of different views \cite{bass2013}: some of them can be described informally or in a purely pictorial way while others are based on a formal model.
In any case, working at the architectural level is a key success factor for privacy by design because architectures carry ``the earliest and hence most fundamental hardest-to-change design decisions''\cite{bass2013}. They should also play a key role to support accountability requirements because they can provide evidence that appropriate decisions have been taken and the reasons for taking them.

Obviously the design of an architecture meeting all privacy requirements is not the end of the story: because an architecture is set, by definition, at a fairly high level of abstraction, the remaining task is to use or devise appropriate mechanisms to implement it. In some sense, the architecture defines precise requirements on the techniques (PETs) to be used in the implementation. The approach presented here is therefore complementary to the work done on the improvement of the implementation of privacy mechanisms \cite{kerschbaum2012}. It is also complementary to  the ongoing work on the conception of new PETs. In practice, a design environment should include a library of available PETs with their associated guarantees (in the privacy logic) and this library should follow  the progress of the technologies. One challenge to this respect will be to ensure that the logical framework is general and versatile enough to allow the description of future mechanisms. We believe that epistemic logics provide a well-established and suitable framework to this aim but this claim has to be confirmed in practice. 

As far as the formal framework itself is concerned, we have suggested a ``logical'' (or qualitative) approach in this paper. An avenue for further research in this area would be to study the integration of quantitative measures of privacy (such as differential privacy) into the framework. This extension would be required to deal with the (numerous) situations in which data cannot be classified into two categories (can or cannot be disclosed) but can be disclosed provided a sufficiently robust sanitization algorithm is applied. Further work on this issue will benefit from existing results~\cite{Dwork:11,McSherry:10,McSherry:07} on the design of mechanisms achieving differential privacy.

In this paper we provided some hints on how our approach can turn into practice based on ongoing work on a privacy by design environment. Needless to say, more work has to be done on the HCI front, to improve the interactions with designers. As suggested in Section 3, graphical means can be used to convey the essence of the properties in the logic and the architecture under construction, for example to picture  data flows, the locations of the computations and the trust relationships. A key concept to this respect is the notion of view in software architectures. More experience is needed, however, to understand what will be the most useful views for a designer. 

Last but not least, another interesting avenue for further research would be to apply this approach to other aspects of privacy (using combinations of formal and semi-formal methods) and to establish a link with the coarse-grain strategies defined in \cite{hoepman2013} to drive the interactions with the system.

\subsubsection{Acknowledgement.}

This work was partially funded by the European project PRIPARE / FP7-ICT-2013-1.5, the ANR project BIOPRIV, and the Inria Project Lab CAPPRIS (Collaborative Action on the Protection of Privacy Rights in the Information Society.)

%
%

\newpage

\section*{\centerline{Annex 1}}

\begin{figure}[H]
\begin{center}
\includegraphics[width=0.98\textwidth]{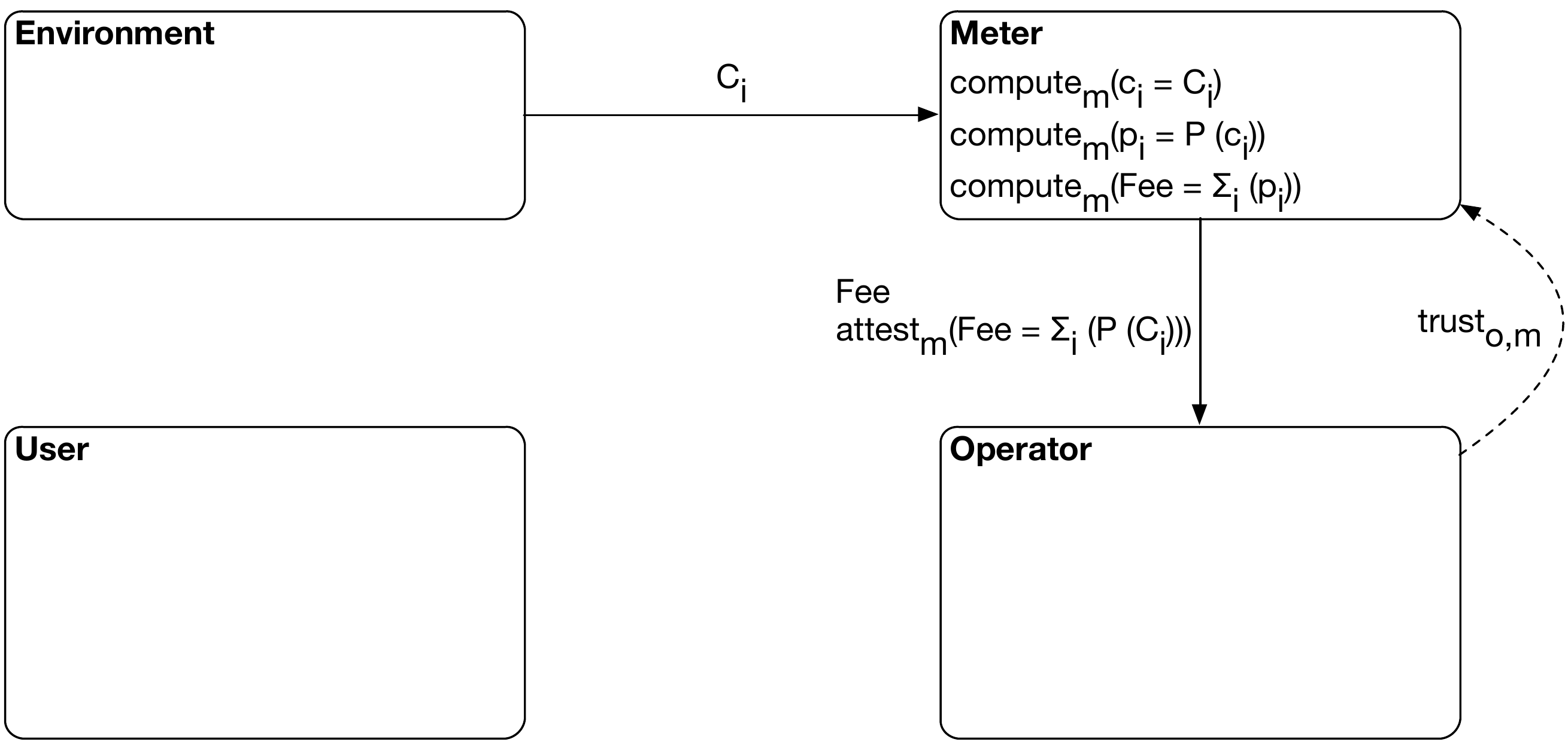}
\end{center}
\caption{Option 2 of Section \ref{sec:example-of-application}: The meter computes the fee and has just to be trusted by the operator. $\mathsf{C_i}$ are the actual consumptions and $\mathsf{c_i}$ the values actually used by the meter.}
\label{sd_option_3}
\end{figure}

\begin{figure}[H]
\begin{center}
\includegraphics[width=0.98\textwidth]{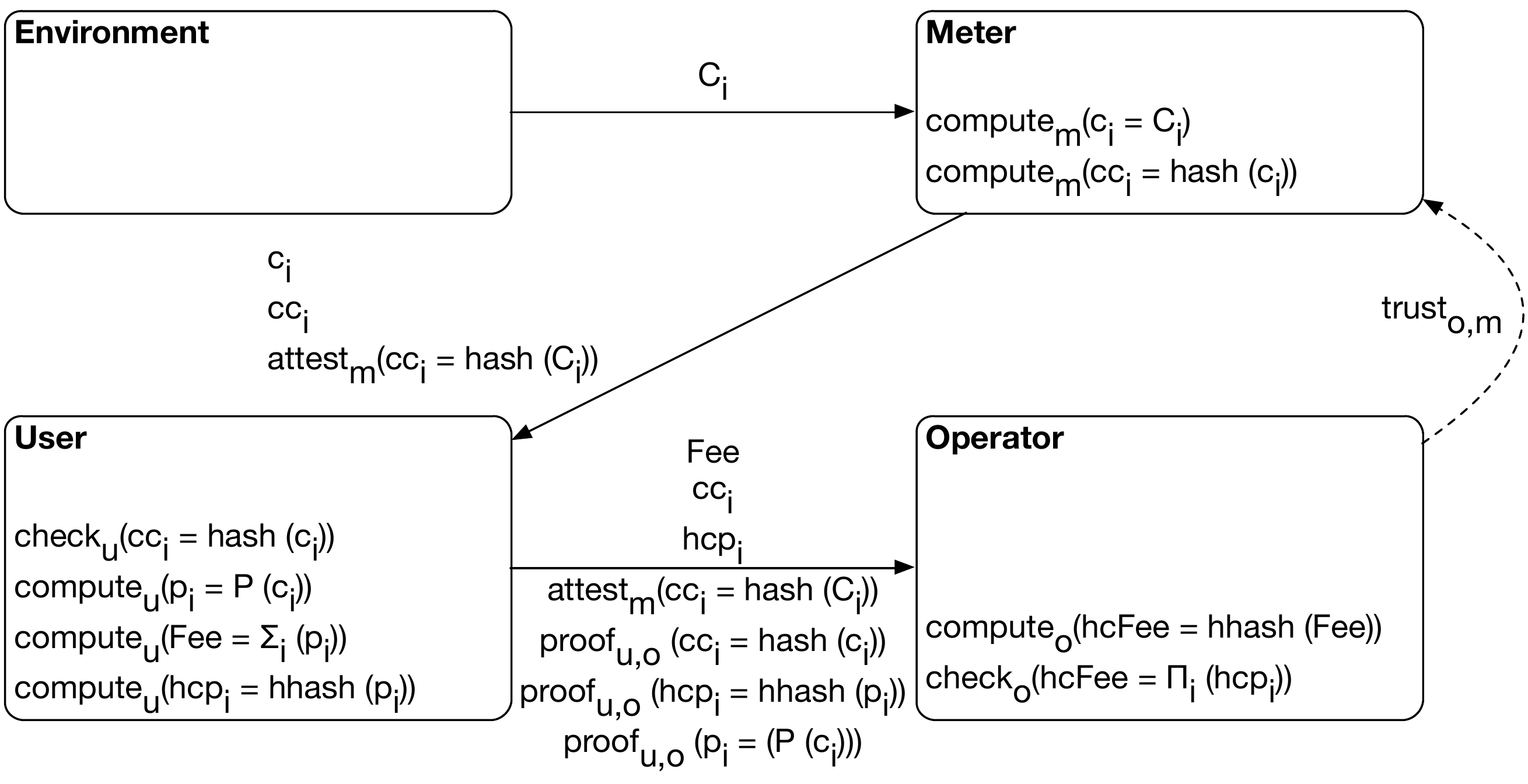}
\end{center}
\caption{Option 3 of Section \ref{sec:example-of-application}: The user computes the fee and the operator has to trust the meter (for using the right value $\mathsf{c_i}$). $\mathsf{hhash}$ is a homomorphic hash function which allows the operator to check that the hash of the global fee $\mathsf{Fee}$ is consistent with the hashes $\mathsf{hcp_i}$ of individual fees $\mathsf{p_i}$. This architecture is an abstraction of the solution proposed by~\cite{Rial}.}
\label{sd_option_1}
\end{figure}

\end{document}